\documentclass[11pt]{article}
\usepackage[letterpaper,hmargin=1in,vmargin=1.25in]{geometry}

\usepackage{url}

\usepackage{amsmath}
\usepackage{amssymb}

\usepackage{amsthm}

\theoremstyle{plain}
\newtheorem{theorem}{Theorem}[section]
\newtheorem{lemma}[theorem]{Lemma}
\newtheorem{corollary}[theorem]{Corollary}

\newtheorem{definition}[theorem]{Definition}

\newtheorem{property}{Property}

\newcommand{\abs}[1]{\left\lvert#1\right\rvert}
\DeclareMathOperator{\Sim}{sim}
\DeclareMathOperator{\Dom}{dom}
\DeclareMathOperator*{\Limsup}{\varlimsup}

\newcommand{\N}{\mathbb{N}}
\newcommand{\Q}{\mathbb{Q}}
\newcommand{\R}{\mathbb{R}}
\newcommand{\X}{\{0,1\}^*}
\newcommand{\K}{H}

\newcommand{\noi}{\noindent}

\begin{document}


\begin{center}
{\Large \textbf{
  Equivalent characterizations of partial randomness\\
  for a recursively enumerable real
}}
\end{center}

\vspace{-2mm}

\begin{center}
Kohtaro Tadaki
\end{center}

\vspace{-5mm}

\begin{center}
Research and Development Initiative, Chuo University\\
1--13--27 Kasuga, Bunkyo-ku, Tokyo 112-8551, Japan\\
E-mail: tadaki@kc.chuo-u.ac.jp
\end{center}

\vspace{-2mm}

\begin{quotation}
\noi\textbf{Abstract.}
A real number $\alpha$ is called recursively enumerable if
there exists a computable,
increasing sequence of rational numbers which converges to $\alpha$.
The randomness of a recursively enumerable real
$\alpha$
can be characterized in various ways
using each of the notions;
program-size complexity,
Martin-L\"{o}f test,
Chaitin's $\Omega$ number,
the domination and $\Omega$-likeness of $\alpha$,
the universality of
a computable, increasing sequence of rational numbers
which converges to $\alpha$,
and universal probability.
In this paper,
we generalize these characterizations of randomness
over the notion of partial randomness
by parameterizing each of the notions above
by a real number $T\in(0,1]$.
We thus present several equivalent characterizations of
partial randomness for a recursively enumerable real number.
\end{quotation}

\begin{quotation}
\noi\textit{Key words\/}:
algorithmic randomness,
recursively enumerable real number,
partial randomness,
Chaitin's $\Omega$ number,
program-size complexity,
universal probability
\end{quotation}

\section{Introduction}

A real number $\alpha$ is called \textit{recursively enumerable}
(``r.e.'' for short)
if there exists a computable, increasing sequence of rational numbers
which converges to $\alpha$.
The randomness of an r.e.~real $\alpha$
can be characterized in various ways
using each of the notions;
\textit{program-size complexity},
\textit{Martin-L\"{o}f test},
\textit{Chaitin's $\Omega$ number},
the \textit{domination} and \textit{$\Omega$-likeness}
of $\alpha$,
the \textit{universality} of
a computable, increasing sequence of rational numbers
which converges to $\alpha$,
and \textit{universal probability}.
These equivalent characterizations of randomness for an r.e.~real number
are summarized in Theorem \ref{randomness}
(see Section \ref{previous results}),
where the equivalences are
established by a series of works of
Schnorr \cite{Sch73},
Chaitin \cite{C75},
Solovay \cite{Sol75},
Calude, Hertling, Khoussainov and Wang \cite{CHKW01},
Ku\v{c}era and Slaman \cite{KS01},
and Tadaki \cite{T06}.
In this paper,
we generalize
these characterizations of randomness
over the notion of \textit{partial randomness},
which was introduced by Tadaki \cite{T99,T02}.
We introduce several characterizations of
partial randomness for an r.e.~real number
by parameterizing each of the notions above on randomness
by a real number $T\in(0,1]$.
We prove the equivalence of all these characterizations
of partial randomness in Theorem \ref{partial randomness},
our main result,
in Section \ref{Our results}.

The paper is organized as follows.
We begin in Section \ref{preliminaries} with
some preliminaries to
algorithmic information theory and partial randomness.
In Section \ref{previous results},
we review the previous results on
the equivalent characterizations of randomness for an r.e.~real number.
Our main result on partial randomness of an r.e.~real number is
presented in Section \ref{Our results},
and its proof is completed in Section \ref{proofs}.
In Section \ref{T-convergence},
we investigate some properties of the notion of \textit{$T$-convergence}
for an increasing sequence of real numbers,
which plays a crucial role in our characterizations of
partial randomness.
We conclude this paper with
a mention of
the future direction of this work in Section \ref{conclusion}.

\section{Preliminaries}
\label{preliminaries}

\subsection{Basic notation}
\label{basic notation}

We start with some notation about numbers and strings
which will be used in this paper.
$\#S$ is the cardinality of $S$ for any set $S$.
$\N=\left\{0,1,2,3,\dotsc\right\}$ is the set of natural numbers,
and $\N^+$ is the set of positive integers.
$\Q$ is the set of rational numbers, and
$\R$ is the set of real numbers.
A sequence $\{a_n\}_{n\in\N}$ of numbers
(rational numbers or real numbers)
is called \textit{increasing} if $a_{n+1}>a_{n}$ for all $n\in\N$.

$\X=
\left\{
  \lambda,0,1,00,01,10,11,000,001,010,\dotsc
\right\}$
is the set of finite binary strings
where $\lambda$ denotes the \textit{empty string},
and $\X$ is ordered as indicated.
We identify any string in $\X$ with a natural number in this order,
i.e.,
we consider $\varphi\colon \X\to\N$ such that $\varphi(s)=1s-1$
where the concatenation $1s$ of strings $1$ and $s$ is regarded
as a dyadic integer,
and then we identify $s$ with $\varphi(s)$.
For any $s \in \X$, $\abs{s}$ is the \textit{length} of $s$.
A subset $S$ of $\X$ is called a \textit{prefix-free set}
if no string in $S$ is a prefix of another string in $S$.
For any partial function $f$,
the domain of definition of $f$ is denoted by $\Dom f$.
We write ``r.e.'' instead of ``recursively enumerable.''

Normally, $o(n)$ denotes any
function $f\colon \N^+\to\R$ such
that $\lim_{n \to \infty}f(n)/n=0$.
On the other hand,
$O(1)$ denotes any
function $g\colon \N^+\to\R$ such that
there is $C\in\R$ with the property that
$\abs{g(n)}\le C$ for all $n\in\N^+$.

Let $\alpha$ be an arbitrary real number.
We denote $\alpha - \lfloor \alpha \rfloor$ by $\alpha\bmod 1$,
where $\lfloor \alpha \rfloor$ is the greatest integer
less than or equal to $\alpha$.
Hence, $\alpha\bmod 1 \in [0,1)$.
Normally, $\lceil \alpha \rceil$ denotes
the smallest integer greater than or equal to $\alpha$.
We denote by $\alpha_n\in\X$
the first $n$ bits of the base-two expansion of $\alpha\bmod 1$
with infinitely many zeros.
Thus,
in particular,
if $\alpha\in[0,1)$,
then $\alpha_n$ denotes the first $n$ bits of
the base-two expansion of $\alpha$ with infinitely many zeros.
For example,
in the case of $\alpha=5/8$,
$\alpha_6=101000$.

A real number $\alpha$ is called \textit{r.e.}~if
there exists a computable,
increasing sequence of rational numbers which converges to $\alpha$.
An r.e.~real number is also called a
\textit{left-computable} real number.
On the other hand,
a real number $\alpha$ is called \textit{right-computable} if
$-\alpha$ is left-computable.
We say that a real number $\alpha$ is \textit{computable} if
there exists a computable sequence $\{a_n\}_{n\in\N}$ of rational numbers
such that $\abs{\alpha-a_n} < 2^{-n}$ for all $n\in\N$.
It is then easy to see that,
for every $\alpha\in\R$,
$\alpha$ is computable if and only if
$\alpha$ is both left-computable and right-computable.
A sequence $\{a_n\}_{n\in\N}$ of real numbers is called
\textit{computable} if
there exists a total recursive function $f\colon\N\times\N \to \Q$
such that
$\abs{a_n-f(n,m)} < 2^{-m}$ for all $n, m\in\N$.
See e.g.~Pour-El and Richards \cite{PR89}
and Weihrauch \cite{W00}
for the detail of the treatment of
the computability of real numbers and
sequences of real numbers.

\subsection{Algorithmic information theory}
\label{ait}

In the following
we concisely review some definitions and results of
algorithmic information theory
\cite{C75,C87b}.
A \textit{computer} is a partial recursive function
$C\colon \X\to \X$
such that
$\Dom C$ is a prefix-free set.
For each computer $C$ and each $s \in \X$,
$\K_C(s)$ is defined by
$\K_C(s) =
\min
\left\{\,
  \abs{p}\,\big|\;p \in \X\>\&\>C(p)=s
\,\right\}$.
A computer $U$ is said to be \textit{optimal} if
for each computer $C$ there exists a constant $\Sim(C)$
with the following property;
if $C(p)$ is defined, then there is a $p'$ for which
$U(p')=C(p)$ and $\abs{p'}\le\abs{p}+\Sim(C)$.
It is easy to see that there exists an optimal computer.
We choose a particular optimal computer $U$
as the standard one for use,
and define $\K(s)$ as $\K_U(s)$,
which is referred to as
the \textit{program-size complexity} of $s$,
the \textit{information content} of $s$, or
the \textit{Kolmogorov complexity} of $s$
\cite{G74,L74,C75}.
Thus, $H(s) \le H_C(s) + \Sim(C)$ for every computer $C$.

Let $V$ be an arbitrary optimal computer.
For each $s\in \X$, $P_V(s)$ is defined as $\sum_{V(p)=s}2^{-\abs{p}}$.
\textit{Chaitin's halting probability} $\Omega_V$ of $V$ is defined by
\begin{equation*}
  \Omega_V=\sum_{p\in\Dom V}2^{-\abs{p}}.
\end{equation*}
Thus, $\Omega_V=\sum_{s\in\X}P_V(s)$.

\begin{definition}[weak Chaitin randomness, Chaitin \cite{C75,C87b}]
  For any $\alpha\in\R$,
  we say that $\alpha$ is \textit{weakly Chaitin random} if
  there exists $c\in\N$ such that
  $n-c \le H(\alpha_n)$
  for all $n\in\N^+$.
  \qed
\end{definition}

Chaitin \cite{C75} showed that,
for every optimal computer $V$,
$\Omega_V$ is weakly Chaitin random.

\begin{definition}[Martin-L\"{o}f randomness, Martin-L\"{o}f \cite{M66}]
A subset $\mathcal{C}$ of $\N^+\times\X$
is called a \textit{Martin-L\"{o}f test} if
$\mathcal{C}$ is an r.e.~set and
\begin{equation*}
  \forall\,n\in\N^+\quad
  \sum_{s\,\in\,\mathcal{C}_n}2^{-\abs{s}}
  \le 2^{-n},
\end{equation*}
where
$\mathcal{C}_n
=
\left\{\,
  s\bigm|(n,s)\in\mathcal{C}
\,\right\}$.
For any $\alpha\in\R$,
we say that $\alpha$ is \textit{Martin-L\"{o}f random} if
for every Martin-L\"{o}f test $\mathcal{C}$,
there exists $n\in\N^+$ such that, for every $k\in\N^+$,
$\alpha_k\notin\mathcal{C}_n$.\qed
\end{definition}

\begin{theorem}[Schnorr \cite{Sch73}]\label{wcem}
  For every $\alpha\in\R$,
  $\alpha$ is weakly Chaitin random if and only if
  $\alpha$ is Martin-L\"{o}f random.\qed
\end{theorem}

It follows from Theorem \ref{wcem} that
$\Omega_V$ is Martin-L\"{o}f random
for every optimal computer $V$.

The program-size complexity $\K(s)$ is originally defined
using the concept of program-size, as stated above.
However,
it is possible to define $\K(s)$ without referring to such a concept,
i.e.,
as in the following,
we first introduce a \textit{universal probability} $m$,
and then define $\K(s)$ as $-\log_2 m(s)$.
A universal probability is defined as follows \cite{ZL70}.

\begin{definition}[universal probability]
  A function $r\colon \X\to[0,1]$ is called
  a \textit{lower-computable semi-measure} if
  $\sum_{s\in \X}r(s)\le 1$ and
  the set $\{(a,s)\in\Q\times\X\mid a<r(s)\}$ is r.e.
  We say that a lower-computable semi-measure $m$ is
  a \textit{universal probability} if
  for every lower-computable semi-measure $r$,
  there exists $c\in\N^+$ such that,
  for all $s\in \X$, $r(s)\le cm(s)$.\qed
\end{definition}

The following theorem can be then shown
(see e.g.~Theorem 3.4 of Chaitin \cite{C75} for its proof).

\begin{theorem}%
\label{eup}
  For every optimal computer $V$,
  both $2^{-\K_V(s)}$ and $P_V(s)$ are universal probabilities.
  \qed
\end{theorem}

By Theorem \ref{eup}, we see that
$\K(s)=-\log_2 m(s)+O(1)$
for every universal probability $m$.
Thus it is possible to define $\K(s)$ as $-\log_2 m(s)$
with a particular universal probability $m$
instead of as $\K_U(s)$.
Note that
the difference up to an additive constant is
nonessential
to algorithmic information theory.
Any universal probability is not computable,
as corresponds to the uncomputability of $\K(s)$.
As a result, we see that
$0<\sum_{s\in\X}m(s)<1$ for every universal probability $m$.

\subsection{Partial randomness}
\label{partial}

In the works \cite{T99,T02},
we generalized the notion of
the randomness of
a real number
so that \textit{the degree of the randomness},
which is often referred to
as
\textit{the partial randomness} recently
\cite{CST06,RS05,CS06},
can be characterized by a real number $T$
with $0<T\le 1$ as follows.

\begin{definition}[weak Chaitin $T$-randomness]
  Let $T\in\R$ with $T\ge 0$.
  For any $\alpha\in\R$,
  we say that $\alpha$ is \textit{weakly Chaitin $T$-random} if
  there exists $c\in\N$ such that
  $Tn-c \le H(\alpha_n)$
  for all $n\in\N^+$.
  \qed
\end{definition}

\begin{definition}[Martin-L\"{o}f $T$-randomness]
Let $T\in\R$ with $T\ge 0$.
A subset $\mathcal{C}$ of $\N^+\times\X$
is called a \textit{Martin-L\"{o}f $T$-test} if
$\mathcal{C}$ is an r.e.~set and
\begin{equation*}
  \forall\,n\in\N^+\quad
  \sum_{s\,\in\,\mathcal{C}_n}2^{-T\abs{s}}
  \le 2^{-n}.
\end{equation*}
For any $\alpha\in\R$,
we say that $\alpha$ is \textit{Martin-L\"{o}f $T$-random} if
for every Martin-L\"{o}f $T$-test $\mathcal{C}$,
there exists $n\in\N^+$ such that, for every $k\in\N^+$,
$\alpha_k\notin\mathcal{C}_n$.\qed
\end{definition}

In the case where $T=1$,
the weak Chaitin $T$-randomness and Martin-L\"{o}f $T$-randomness
result in weak Chaitin randomness and Martin-L\"{o}f randomness,
respectively.
Tadaki \cite{T02} generalized Theorem \ref{wcem}
over the notion of $T$-randomness
as follows.

\begin{theorem}[Tadaki \cite{T02}]\label{Twcem}
  Let $T$ be a computable real number with $T\ge 0$.
  Then,
  for every $\alpha\in\R$,
  $\alpha$ is weakly Chaitin $T$-random if and only if
  $\alpha$ is Martin-L\"{o}f $T$-random.\qed
\end{theorem}

\begin{definition}[$T$-compressibility]
Let $T\in\R$ with $T\ge 0$.
For any $\alpha\in\R$,
we say that $\alpha$ is \textit{$T$-compressible} if
$H(\alpha_n)\le Tn+o(n)$,
which is equivalent to
\begin{equation*}
  \Limsup_{n \to \infty}\frac{H(\alpha_n)}{n}\le T.
\end{equation*}
\qed
\end{definition}

For every $T\in[0,1]$ and every $\alpha\in\R$,
if $\alpha$ is weakly Chaitin $T$-random and $T$-compressible,
then
\begin{equation}\label{compression-rate}
  \lim_{n\to \infty} \frac{H(\alpha_n)}{n} = T,
\end{equation}
and therefore
the \textit{compression rate} of $\alpha$
by the program-size complexity $H$ is equal to $T$.
Note, however, that \eqref{compression-rate}
does not necessarily imply that $\alpha$ is weakly Chaitin $T$-random.

In the works \cite{T99,T02},
we generalized Chaitin's halting probability $\Omega$ to $\Omega(T)$
as follows.
For each optimal computer $V$ and each real number $T>0$,
the \textit{generalized halting probability} $\Omega_V(T)$ of $V$ is
defined by
\begin{equation*}
  \Omega_V(T) = \sum_{p\in\Dom V}2^{-\frac{\abs{p}}{T}}.
\end{equation*}
Thus,
$\Omega_V(1)=\Omega_V$.
If $0<T\le 1$, then $\Omega_V(T)$ converges and $0<\Omega_V(T)<1$,
since $\Omega_V(T)\le \Omega_V<1$.
The following theorem holds for $\Omega_V(T)$.

\begin{theorem}[Tadaki \cite{T99,T02}]\label{pomgd}
Let $V$ be an optimal computer and let $T\in\R$.
\begin{enumerate}
  \item If $0<T\le 1$ and $T$ is computable,
    then $\Omega_V(T)$ is weakly Chaitin $T$-random and
    $T$-compressible.
  \item If $1<T$, then $\Omega_V(T)$ diverges to $\infty$.\qed
\end{enumerate}
\end{theorem}

Note also that
the computability of $\Omega_V(T)$ gives
a sufficient condition for a real number $T\in(0,1)$
to be a \textit{fixed point on partial randomness}
as follows.

\begin{theorem}[Tadaki \cite{T08CiE}]
Let $V$ be an optimal computer.
For every $T\in(0,1)$,
if $\Omega_V(T)$ is computable,
then $T$ is weakly Chaitin $T$-random and $T$-compressible,
and therefore
\begin{equation*}
  \lim_{n\to\infty}\frac{H(T_n)}{n}=T.
\end{equation*}
\qed
\end{theorem}

\section{Previous results on the randomness of an r.e.~real}
\label{previous results}

In this section,
we review the previous results on the randomness of an r.e.~real number.
First we review some notions on r.e.~real numbers.

\begin{definition}[$\Omega$-likeness]
For any r.e.~real numbers $\alpha$ and $\beta$,
we say that $\alpha$ \textit{dominates} $\beta$
if there are computable, increasing sequences $\{a_n\}$ and $\{b_n\}$
of rational numbers and $c\in\N^+$
such that
$\lim_{n\to\infty} a_n =\alpha$,
$\lim_{n\to\infty} b_n =\beta$,
and $c(\alpha-a_n)\ge \beta-b_n$ for all $n\in\N$.
An r.e.~real number $\alpha$ is called \textit{$\Omega$-like}
if it dominates all r.e.~real numbers.
\qed
\end{definition}

Solovay
\cite{Sol75}
showed the following theorem.
For
its
proof, see also Theorem 4.9 of \cite{CHKW01}.

\begin{theorem}[Solovay \cite{Sol75}]\label{Solovay}
For every r.e.~real numbers $\alpha$ and $\beta$,
if $\alpha$ dominates $\beta$ then
$H(\beta_n)\le H(\alpha_n)+O(1)$.
\qed
\end{theorem}

\begin{definition}[universality]
A computable, increasing and converging sequence $\{a_n\}$
of rational numbers is called \textit{universal}
if for every computable, increasing and converging sequence $\{b_n\}$
of rational numbers
there exists $c\in\N^+$ such that
$c(\alpha-a_n)\ge \beta-b_n$ for all $n\in\N$,
where $\alpha=\lim_{n\to\infty} a_n$ and $\beta=\lim_{n\to\infty} b_n$.
\qed
\end{definition}

The previous results on
the equivalent characterizations of randomness for an r.e.~real number
are summarized in the following theorem.

\begin{theorem}[\cite{Sch73,C75,Sol75,CHKW01,KS01,T06}]\label{randomness}
Let $\alpha$ be an r.e.~real number with $0<\alpha<1$.
Then the following conditions are equivalent:
\begin{enumerate}
\item The real number $\alpha$ is weakly Chaitin random.
\item The real number $\alpha$ is Martin-L\"{o}f random.
\item The real number $\alpha$ is $\Omega$-like.
\item $H(\beta_n)\le H(\alpha_n)+O(1)$ for every r.e.~real number $\beta$.
\item There exists an optimal computer $V$ such that $\alpha=\Omega_V$.
\item There exists a universal probability $m$ such that
  $\alpha=\sum_{s\in\X}m(s)$.
\item Every computable, increasing sequence of rational numbers
  which converges to $\alpha$ is universal.
\item There exists a universal computable, increasing sequence of
  rational numbers which converges to $\alpha$.\qed
\end{enumerate}
\end{theorem}

The historical remark on the proofs of equivalences
in Theorem \ref{randomness} is as follows.
Schnorr \cite{Sch73} showed that
(i) and (ii) are equivalent to each other.
Chaitin \cite{C75} showed that (v) implies (i).
Solovay \cite{Sol75} showed that
(v) implies (iii), (iii) implies (iv), and (iii) implies (i).
Calude, Hertling, Khoussainov, and Wang \cite{CHKW01} showed that
(iii) implies (v), and (v) implies (vii).
Ku\v{c}era and Slaman \cite{KS01} showed that
(ii) implies (vii).
Finally,
(vi) was inserted in the course of the derivation from (v) to (viii)
by Tadaki \cite{T06}.

\section{New results on the partial randomness of an r.e.~real}
\label{Our results}

In this section,
we generalize
Theorem \ref{randomness} above
over the notion of partial randomness.
For that purpose,
we first introduce some new notions.
Let $T$ be an arbitrary
real number with $0<T\le 1$
throughout
the rest of this paper.
These notions are parametrized by the real number $T$.

\begin{definition}[$T$-convergence]
An increasing sequence $\{a_n\}$ of real numbers is called
\textit{$T$-convergent} if
$\sum_{n=0}^{\infty} (a_{n+1}-a_{n})^T<\infty$.
An r.e.~real number $\alpha$ is called \textit{$T$-convergent} if
there exists a $T$-convergent computable,
increasing sequence of rational numbers which
converges to $\alpha$.
\qed
\end{definition}

Note that
every increasing and converging sequence of real numbers is
$1$-convergent, and thus
every r.e.~real number is $1$-convergent.
In general,
based on the following
lemma,
we can freely switch from
``$T$-convergent computable,
increasing sequence of real numbers''
to
``$T$-convergent computable,
increasing sequence of rational numbers.''

\begin{lemma}\label{tcrr}
For every $\alpha\in\R$,
$\alpha$ is an r.e.~$T$-convergent real number
if and only if
there exists a $T$-convergent computable,
increasing sequence of real numbers which
converges to $\alpha$.
\end{lemma}

\begin{proof}
The ``only if'' part is obvious.
We show the ``if'' part.
Suppose that
$\{a_n\}$ is a $T$-convergent computable,
increasing sequence of real numbers which
converges to $\alpha$.
Then, we first see that
there exists a computable sequence $\{b_n\}$ of rational numbers
such that
$a_{n}<b_{n}<a_{n+1}$
for all $n\in\N$.
Obviously,
$\{b_n\}$ is an increasing sequence of rational numbers which
converges to $\alpha$.
On the other hand,
using the inequality $(x+y)^t\le x^t+y^t$
for real numbers $x,y>0$ and $t\in(0,1]$,
we see that
$(b_{n+1}-b_{n})^T<(a_{n+2}-a_{n})^T
\le(a_{n+2}-a_{n+1})^T+(a_{n+1}-a_{n})^T$.
Thus,
since
$\sum_{n=0}^{\infty} (a_{n+2}-a_{n+1})^T$ and
$\sum_{n=0}^{\infty} (a_{n+1}-a_{n})^T$ both converge,
the increasing sequence $\{b_n\}$
of rational numbers
is $T$-convergent.
\end{proof}

The following argument illustrates the way of using
Lemma
\ref{tcrr}:
Let $V$ be an optimal computer, and
let $p_0,p_1,p_2,\dots$ be a recursive enumeration of the r.e.~set $\Dom V$.
Then $\Omega_V(T)=\sum_{i=0}^{\infty} 2^{-\abs{p_i}/T}$,
and the increasing sequence
$\left\{\sum_{i=0}^{n} 2^{-\abs{p_i}/T} \right\}_{n\in\N}$
of real numbers is $T$-convergent
since
$\Omega_V=\sum_{i=0}^{\infty} 2^{-\abs{p_i}}<1$.
If $T$ is computable,
then this sequence of real numbers is
computable. 
Thus,
by Lemma \ref{tcrr}
we have Theorem \ref{tomegavt} below.

\begin{theorem}\label{tomegavt}
Let $V$ be an optimal computer.
If $T$ is computable,
then $\Omega_V(T)$ is an r.e.~$T$-convergent real number.
\qed
\end{theorem}

\begin{definition}[$\Omega(T)$-likeness]
An r.e.~real number $\alpha$ is called \textit{$\Omega(T)$-like}
if it dominates all r.e.~$T$-convergent real numbers.
\qed
\end{definition}

Note that
an r.e.~real number $\alpha$ is $\Omega(1)$-like
if and only if $\alpha$ is $\Omega$-like.

\begin{definition}[$T$-universality]
A computable, increasing and converging sequence $\{a_n\}$
of rational numbers is called \textit{$T$-universal}
if for every $T$-convergent computable,
increasing and converging sequence $\{b_n\}$ of
rational
numbers
there exists $c\in\N^+$ such that
$c(\alpha-a_n)\ge \beta-b_n$ for all $n\in\N$,
where $\alpha=\lim_{n\to\infty} a_n$ and $\beta=\lim_{n\to\infty} b_n$.
\qed
\end{definition}

Note that
a computable, increasing and converging sequence $\{a_n\}$
of rational numbers is $1$-universal
if and only if $\{a_n\}$ is universal.

Using the notions introduced above,
Theorem \ref{randomness} is generalized as follows.

\begin{theorem}[main result]\label{partial randomness}
Let $\alpha$ be an r.e.~real number with $0<\alpha<1$.
Suppose that $T$ is computable.
Then the following conditions are equivalent:
\begin{enumerate}
\item The real number $\alpha$ is weakly Chaitin $T$-random.
\item The real number $\alpha$ is Martin-L\"{o}f $T$-random.
\item The real number $\alpha$ is $\Omega(T)$-like.
\item $H(\beta_n)\le H(\alpha_n)+O(1)$
  for every r.e.~$T$-convergent real number $\beta$.
\item For every r.e.~$T$-convergent real number $\gamma>0$,
  there exist an r.e.~real number $\beta\ge 0$ and
  a rational number $q>0$ such that
  $\alpha=\beta+q\gamma$.
\item There exist
  an optimal computer $V$ and an r.e.~real number $\beta\ge 0$
  such that $\alpha=\beta+\Omega_V(T)$.
\item There exists a universal probability $m$ such that
  $\alpha=\sum_{s\in\X}m(s)^{\frac{1}{T}}$.
\item Every computable, increasing sequence of rational numbers
  which converges to $\alpha$ is $T$-universal.
\item There exists a $T$-universal computable, increasing sequence
  of rational numbers which converges to $\alpha$.\qed
\end{enumerate}
\end{theorem}

The condition (vi) of Theorem \ref{partial randomness}
corresponds to the condition (v) of Theorem \ref{randomness}.
Note, however, that,
in the condition (vi) of Theorem \ref{partial randomness},
a non-negative r.e.~real number $\beta$ is needed.
The reason is as follows:
In the case of $\beta=0$,
the possibility that $\alpha$ is weakly Chaitin $T'$-random
with a real number $T'>T$
is excluded by
the $T$-compressibility of $\Omega_V(T)$
imposed by Theorem \ref{pomgd} (i).
However, this exclusion is inconsistent with
the condition (i) of Theorem \ref{partial randomness}.

Theorem \ref{partial randomness} is proved as follows,
partially based on
Theorems
\ref{(ii)->(v)}, \ref{(v)->(vi)},
\ref{(v)->(vii)}, and \ref{(vii)->(viii)},
which
will be proved in the next section.

\begin{proof}[Proof of Theorem \ref{partial randomness}]
We prove the equivalences in Theorem \ref{partial randomness}
by showing
the two paths [A] and [B] of implications
below.

[A] The
implications
(i) $\Rightarrow$ (ii)
$\Rightarrow$ (v)
$\Rightarrow$ (vi)
$\Rightarrow$ (i):
First,
by Theorem \ref{Twcem}, (i) implies (ii) obviously.
It follows from Theorem \ref{(ii)->(v)} below
that (ii) implies (v),
and also
it follows from Theorem \ref{(v)->(vi)} below
that (v) implies (vi).
For the forth implication,
let $V$ be an optimal computer,
and let $\beta$ be an r.e.~real number.
It is then easy to show that
$\beta+\Omega_V(T)$ dominates $\Omega_V(T)$
(see the condition 2 of Lemma 4.4 of \cite{CHKW01}).
It follows from
Theorem \ref{Solovay} and Theorem \ref{pomgd} (i)
that
the condition (vi) results in the condition (i)
of Theorem \ref{partial randomness}.

[B] The
implications
(v) $\Rightarrow$ (vii)
$\Rightarrow$ (viii)
$\Rightarrow$ (ix)
$\Rightarrow$ (iii)
$\Rightarrow$ (iv)
$\Rightarrow$ (i):
First,
it follows from Theorem \ref{(v)->(vii)} below
that (v) implies (vii),
and also it follows from Theorem \ref{(vii)->(viii)} below
that (vii) implies (viii).
Obviously,
(viii) implies (ix) and (ix) implies (iii).
It follows from Theorem \ref{Solovay} that (iii) implies (iv).
Finally,
note that
$\Omega_U(T)$ is an r.e.~$T$-convergent real number
which is weakly Chaitin $T$-random
by Theorem \ref{pomgd} (i) and Theorem \ref{tomegavt}.
Thus, by setting $\beta$ to $\Omega_U(T)$ in the condition (iv),
the condition (iv) results in the condition (i).
\end{proof}

As a consequence of Theorem \ref{partial randomness},
we obtain the following corollary, for example.

\begin{corollary}\label{omegat-omegat}
Suppose that $T$ is computable.
Then, for every two optimal computers $V$ and $W$,
\begin{equation*}
  H((\Omega_{V}(T))_n)=H((\Omega_{W}(T))_n)+O(1).
\end{equation*}
\end{corollary}

\begin{proof}
Corollary \ref{omegat-omegat}
follows immediately from Theorem \ref{tomegavt} and
the implication (vi) $\Rightarrow$ (iv) of
Theorem \ref{partial randomness}.
\end{proof}

\section{The completion of the proof of the main result}
\label{proofs}

In this section,
we prove several theorems needed to complete
the proof of Theorem \ref{partial randomness}.
For the sake of convenience,
we first rephrase the definition of Martin-L\"{o}f $T$-randomness
of a real number as follows.
We denote by $\mathcal{I}$
the set $\{(n,q,r)\in\N^+\times\Q\times\Q\mid q<r\}$.
A subset $\mathcal{D}$ of $\mathcal{I}$
is called a \textit{rational Martin-L\"{o}f $T$-test} if
$\mathcal{D}$ is an r.e.~set and
\begin{equation*}
  \forall\,n\in\N^+\quad
  \sum_{(q,r)\,\in\,\mathcal{D}(n)}(r-q)^T
  \le 2^{-n},
\end{equation*}
where
$\mathcal{D}(n)
=
\left\{\,
  (q,r)\bigm|(n,q,r)\in\mathcal{D}
\,\right\}$.
We can then show the following lemma,
which rephrases the definition of the Martin-L\"{o}f $T$-randomness
of a real number to give it more flexibility.

\begin{lemma}\label{rational-ml}
For every $\alpha\in\R$,
$\alpha$ is Martin-L\"{o}f $T$-random if and only if
for every rational Martin-L\"{o}f $T$-test $\mathcal{D}$,
there exists $n\in\N^+$ such that,
for every $q,r\in\Q$,
if $(q,r)\,\in\,\mathcal{D}(n)$
then $\alpha\notin[q,r]$,
where $[q,r]=\{x\in\R\mid\ q\le x\le r\}$.
\end{lemma}

\begin{proof}
First,
we show the ``if'' part by showing its contraposition.
Suppose that $\alpha$ is not Martin-L\"{o}f $T$-random.
Then
there exists a Martin-L\"{o}f $T$-test $\mathcal{C}$
such that
\begin{equation}\label{nonmlr}
  \forall\,n\in\N^+\;\;
  \exists\,k\in\N^+\;\;
  \alpha_k\in\mathcal{C}_n.
\end{equation}
We define a set $\mathcal{D}\subset \mathcal{I}$ by
\begin{equation*}
  \mathcal{D}
  =
  \{(n,0.s+\lfloor \alpha \rfloor,0.s+2^{-\abs{s}}+\lfloor \alpha \rfloor)
  \mid s\in\mathcal{C}_n\}.
\end{equation*}
Since $\mathcal{C}$ is an r.e.~set,
$\mathcal{D}$ is also an r.e.~set.
We also see that, for each $n\in\N^+$,
\begin{equation*}
  \sum_{(q,r)\,\in\,\mathcal{D}(n)}(r-q)^T
  =
  \sum_{s\,\in\,\mathcal{C}_n}2^{-T\abs{s}}
  \le 2^{-n}.
\end{equation*}
Thus, $\mathcal{D}$ is a rational Martin-L\"{o}f $T$-test.
On the other hand,
note that
$\beta\in
[0.\beta_k+\lfloor \beta \rfloor,
0.\beta_k+2^{-\abs{\beta_k}}+\lfloor \beta \rfloor]$
for every $\beta\in\R$ and every $k\in\N^+$.
It follows from \eqref{nonmlr} that,
for every $n\in\N^+$,
there exist $q,r\in\Q$ such that
$(q,r)\,\in\,\mathcal{D}(n)$ and $\alpha\in[q,r]$.
This completes the proof of the ``if'' part.

Next,
we show the ``only if'' part by showing its contraposition.
Suppose that
there exists a rational Martin-L\"{o}f $T$-test $\mathcal{D}$
such that
\begin{equation}\label{nonrmlr}
  \forall\,n\in\N^+\;\;
  \exists\,q,r\in\Q\;\;
  [\,(q,r)\in\,\mathcal{D}(n)\text{ \& }\alpha\in[q,r]\,].
\end{equation}

In the case of $\alpha\in\Q$,
$\alpha$ is not Martin-L\"{o}f $T$-random, obviously.
This can be shown as follows.
We choose any one $m\in\N^+$ with $Tm\ge 1$.
We then define a set $\mathcal{C}\subset\N^+\times\X$ by
$\mathcal{C}=\{(n,\alpha_{mn})\mid n\in\N^+\}$.
Recall here that $\alpha_{mn}\in\X$ denotes
the first $mn$ bits of the base-two expansion of
$\alpha\bmod 1$ with infinitely many zeros.
Obviously, $\mathcal{C}$ is an r.e.~set.
We also see that, for each $n\in\N^+$,
\begin{equation*}
  \sum_{s\,\in\,\mathcal{C}_n}2^{-T\abs{s}}
  =2^{-Tmn}
  \le 2^{-n}.
\end{equation*}
Therefore, $\mathcal{C}$ is a Martin-L\"{o}f $T$-test.
On the other hand,
$\alpha_{mn}\in\mathcal{C}_n$ for every $n\in\N^+$.
Hence, $\alpha$ is not Martin-L\"{o}f $T$-random,
as desired.

Thus, in what follows, we assume that $\alpha\notin\Q$.
We choose any one $n_0\in\N$ such that
\begin{equation*}
  2^{-\frac{n_0}{T}}
  <\min\{
    \alpha - \lfloor\alpha\rfloor,
    \lfloor \alpha \rfloor+1-\alpha
  \}.
\end{equation*}
We then define a set
$\mathcal{D}^{(0)}\subset \mathcal{I}$
by
\begin{equation*}
  \mathcal{D}^{(0)}
  =
  \{(n,q-\lfloor \alpha \rfloor,r-\lfloor \alpha \rfloor)
  \mid
  n\in\N^+
  \text{ \& }
  (n+n_0,q,r)\in\mathcal{D}
  \text{ \& }
  \lfloor\alpha\rfloor<q,r<\lfloor \alpha \rfloor+1\}.
\end{equation*}
Obviously, $\mathcal{D}^{(0)}$ is an r.e.~set.
We also see that
\begin{equation*}
  \sum_{(q,r)\,\in\,\mathcal{D}^{(0)}(n)}(r-q)^T
  \le
  \sum_{(q,r)\,\in\,\mathcal{D}(n+n_0)}(r-q)^T
  \le 2^{-(n+n_0)}
  \le 2^{-n}
\end{equation*}
for each $n\in\N^+$.
Thus, $\mathcal{D}^{(0)}$ is
a rational Martin-L\"{o}f $T$-test, and also
$\mathcal{D}^{(0)}\subset\N^+\times (0,1)\times (0,1)$.
On the other hand, by the choice of $n_0$,
it is easy to see that,
for every $(n,q,r)\in\mathcal{I}$,
if $(q,r)\in\mathcal{D}(n+n_0)$ and $\alpha\in[q,r]$,
then
$r-q\le 2^{-n_0/T}$ and therefore
$\lfloor\alpha\rfloor<q,r<\lfloor \alpha \rfloor+1$.
It follows from \eqref{nonrmlr} that
\begin{equation}\label{nonrmlr0}
  \forall\,n\in\N^+\;\;
  \exists\,q,r\in\Q\;\;
  [\,(q,r)\in\,\mathcal{D}^{(0)}(n)\text{ \& }
  \alpha\bmod 1\in[q,r]\,].
\end{equation}

For each $q,r\in\Q$ with $0<q<r<1$,
let $v(q,r)$ and $w(q,r)$ be finite binary strings such that
(i) $v(q,r)=q_k$ and $w(q,r)=r_k$ for some $k\in\N^+$, and
(ii) $v(q,r)+1=w(q,r)$
where $v(q,r)$ and $w(q,r)$ are regarded as a dyadic integer.
Such a pair $(v(q,r), w(q,r))$ of finite binary strings
exists uniquely since $0<q<r<1$.
Then,
for every $q,r\in\Q$ with $0<q<r<1$,
it follows that
(i) $2^{-\abs{v(q,r)}}=2^{-\abs{w(q,r)}}\le r-q$, and
(ii) for every $\beta\in\R$,
if $\beta\bmod 1\in[q,r]$ then there exists $k\in\N^+$
such that either $\beta_k=v(q,r)$ or $\beta_k=w(q,r)$.
We define a set $\mathcal{C}\subset\N^+\times\X$ by
\begin{equation*}
  \mathcal{C}
  =
  \bigcup_{(n+1,q,r)\in\mathcal{D}^{(0)}}
  \{(n,v(q,r)),(n,w(q,r))\}.
\end{equation*}
Note that, given $q,r\in\Q$ with $0<q<r<1$,
one can compute both $v(q,r)$ and $w(q,r)$.
Thus, since $\mathcal{D}^{(0)}$ is an r.e.~set,
$\mathcal{C}$ is also an r.e.~set.
We also see that, for each $n\in\N^+$,
\begin{equation*}
  \sum_{s\,\in\,\mathcal{C}_n}2^{-T\abs{s}}
  =
  \sum_{(q,r)\in\mathcal{D}^{(0)}(n+1)}
  \left\{
    2^{-T\abs{v(q,r)}}+2^{-T\abs{w(q,r)}}
  \right\}
  \le
  \sum_{(q,r)\in\mathcal{D}^{(0)}(n+1)}
  2(r-q)^T
  \le 2^{-n}.
\end{equation*}
Thus, $\mathcal{C}$ is a Martin-L\"{o}f $T$-test.
On the other hand,
it is easy to see that,
for every $(n,q,r)\in\mathcal{I}$,
if $(q,r)\in\mathcal{D}^{(0)}(n+1)$
and $\alpha\bmod 1\in[q,r]$,
then
there exists $k\in\N^+$
such that either $\alpha_k=v(q,r)$ or $\alpha_k=w(q,r)$,
and therefore $(n,\alpha_k)\in\mathcal{C}$.
It follows from \eqref{nonrmlr0} that, for every $n\in\N^+$,
there exists $k\in\N^+$ such that $\alpha_k\in\mathcal{C}_n$.
Thus, $\alpha$ is not Martin-L\"{o}f $T$-random.
This completes the proof of the ``only if'' part.
\end{proof}

Lemma \ref{ks} and Theorem \ref{(ii)->(iii)} below
can be proved,
based on the generalization of the techniques used
in the proof of Theorem 2.1 of Ku\v{c}era and Slaman \cite{KS01}
over partial randomness.
We also use Lemma \ref{rational-ml} to prove Lemma \ref{ks} below.

\begin{lemma}\label{ks}
Let $\alpha$ be an r.e.~real number,
and
let $\{d_n\}$ be a computable sequence
of positive rational numbers such that
$\sum_{n=0}^{\infty} {d_n}^T\le 1$.
If $\alpha$ is Martin-L\"{o}f $T$-random,
then for every $\varepsilon>0$
there exist a computable, increasing sequence $\{a_n\}$
of rational numbers and a rational number $q>0$ such that
$a_{n+1}-a_{n}>qd_{n}$ for every $n\in\N$,
$a_{0}>\alpha-\varepsilon$, and
$\alpha=\lim_{n\to\infty}a_{n}$.
\end{lemma}

\begin{proof}
We choose any one rational number $r$ with $2^{-1/T}\ge r>0$.
Since $\alpha$ is an r.e.~real number,
there exists a computable, increasing sequences $\{b_n\}$
of rational numbers such that
$b_{0}>\alpha-\varepsilon$ and
$\alpha=\lim_{n\to\infty}b_{n}$.
We construct a rational Martin-L\"{o}f $T$-test $\mathcal{D}$
by enumerating $\mathcal{D}(i)$ for each $i\in\N^+$ as follows.
During the enumeration of $\mathcal{D}(i)$
we simultaneously construct 
a sequence $\{a(i)_n\}_n$ of rational numbers.

Initially, we set $\mathcal{D}(i):=\emptyset$ and then
specify $a(i)_0$ by $a(i)_0:=b_0$.
In general,
whenever $a(i)_n$ is specified as $a(i)_n:=b_m$,
we update $\mathcal{D}(i)$ by
$\mathcal{D}(i):=\mathcal{D}(i)\cup\{(i,a(i)_n,a(i)_n+r^id_n)\}$,
and calculate $b_{m+1}, b_{m+2}, b_{m+3}, \dotsm$ one by one.
During the calculation,
if we find $m_1$ such that $m_1>m$ and $b_{m_1}>a(i)_n+r^id_n$,
then we specify $a(i)_{n+1}$ by $a(i)_{n+1}:=b_{m_1}$
and we repeat this procedure for
$n+1$.

For the completed
$\mathcal{D}$
through the above procedure,
we see that,
for every $i\in\N^+$,
\begin{equation*}
  \sum_{(r_1,r_2)\,\in\,\mathcal{D}(i)}(r_2-r_1)^T
  =
  \sum_{n} (r^id_n)^T
  \le
  2^{-i}\sum_{n} {d_n}^T
  \le 2^{-i}.
\end{equation*}
Here
the second and third sums
on $n$
may be finite or infinite.
Thus, $\mathcal{D}$ is a rational Martin-L\"{o}f $T$-test.
Since $\alpha$ is Martin-L\"{o}f $T$-random,
there exists $k\in\N^+$ such that,
for every $r_1,r_2\in\Q$,
if $(r_1,r_2)\,\in\,\mathcal{D}(k)$
then $\alpha\notin[r_1,r_2]$.
It follows from $\alpha=\lim_{n\to\infty}b_{n}$ that
in the above procedure for enumerating $\mathcal{D}(k)$,
for every $n\in\N$ we ever find $m_1$ such that
$m_1>m$ and $b_{m_1}>a(k)_n+r^kd_n$.
Therefore,
$\mathcal{D}(k)$ is constructed as an infinite set and also
$\{a(k)_n\}_n$ is constructed as an infinite sequence of
rational numbers.
Thus, we have
$a(k)_{n+1}>a(k)_n+r^kd_n$ for all $n\in\N$.
Since $\{a(k)_n\}_n$ is a subsequence of $\{b_n\}$,
it follows that
the sequence $\{a(k)_n\}_n$ is increasing,
$a(k)_{0}>\alpha-\varepsilon$,
and $\alpha=\lim_{n\to\infty}a(k)_{n}$.
This completes the proof.
\end{proof}

\begin{theorem}\label{(ii)->(v)}
Suppose that $T$ is computable.
For every r.e.~real number $\alpha>0$,
if $\alpha$ is Martin-L\"{o}f $T$-random,
then for every r.e.~$T$-convergent real number $\gamma>0$
there exist an r.e.~real number $\beta>0$ and
a rational number $q>0$ such that
$\alpha=\beta+q\gamma$.
\end{theorem}

\begin{proof}
Suppose that
$\gamma$ is an arbitrary r.e.~$T$-convergent real number
with $\gamma>0$.
Then there exists a $T$-convergent computable,
increasing sequence $\{c_n\}$ of rational numbers which
converges to $\gamma$.
Since $\gamma>0$,
without loss of generality
we can assume that $c_{0}=0$.
We choose any one rational number $\varepsilon>0$
such that
\begin{equation*}
  \sum_{n=0}^{\infty}(c_{n+1}-c_{n})^T
  \le
  \left(
    \frac{1}{\varepsilon}
  \right)^T.
\end{equation*}
Such $\varepsilon$ exists
since the sequence $\{c_n\}$ is $T$-convergent.
It follows that
\begin{equation*}
  \sum_{n=0}^{\infty}
  \left[
    \varepsilon (c_{n+1}-c_{n})
  \right]^T
  \le 1.
\end{equation*}
Note that the sequence $\{\varepsilon (c_{n+1}-c_{n})\}$
is a computable sequence of positive rational numbers.
Thus,
since $\alpha$ is r.e.~and Martin-L\"{o}f $T$-random
by the assumption,
it follows from Lemma \ref{ks}
that there exist a computable, increasing sequence $\{a_n\}$
of rational numbers and a rational number $r>0$ such that
$a_{n+1}-a_{n}>r\varepsilon (c_{n+1}-c_{n})$ for every $n\in\N$,
$a_{0}>0$, and
$\alpha=\lim_{n\to\infty}a_{n}$.
We then define a sequence $\{b_n\}$ of positive real numbers
by $b_n=a_{n+1}-a_{n}-r\varepsilon (c_{n+1}-c_{n})$.
It follows that $\{b_n\}$ is a computable sequence of
rational numbers and
$\sum_{n=0}^{\infty}b_n$ converges to
$\alpha-a_{0}-r\varepsilon (\gamma-c_{0})$.
Thus we have
$\alpha=a_{0}+\sum_{n=0}^{\infty}b_n+r\varepsilon \gamma$,
where $a_{0}+\sum_{n=0}^{\infty}b_n$ is a positive r.e.~real number.
This completes the proof.
\end{proof}

\begin{theorem}\label{(v)->(vi)}
Suppose that $T$ is computable.
For every real number $\alpha$,
if for every r.e.~$T$-convergent real number $\gamma>0$
there exist an r.e.~real number $\beta\ge 0$ and
a rational number $q>0$ such that
$\alpha=\beta+q\gamma$,
then there exist
an optimal computer $V$ and an r.e.~real number $\beta\ge 0$
such that $\alpha=\beta+\Omega_V(T)$.
\end{theorem}

\begin{proof}
First,
for the optimal computer $U$,
it follows from Theorem \ref{tomegavt} that
$\Omega_U(T)$ is an r.e.~$T$-convergent real number.
Thus, by the assumption
there exist an r.e.~real number $\beta\ge 0$ and
a rational number $q>0$ such that
$\alpha=\beta+q\Omega_U(T)$.
We choose any one $n\in\N$ with $q>2^{-n/T}$.
We then define a partial function $V\colon\X\to\X$
by the conditions that
(i)
$\Dom V=\{0^np\mid p\in\Dom U\}$ and
(ii) for every $p\in\Dom U$, $V(0^np)=U(p)$.
Since $\Dom V$ is a prefix-free set,
it follows that $V$ is a computer.
It is then easy to see that
$H_V(s)=H_U(s)+n$ for every $s\in\X$.
Therefore, since $U$ is an optimal computer,
$V$ is also an optimal computer.
It follows that $\Omega_V(T)=2^{-n/T}\Omega_U(T)$.
Thus we have
$\alpha=\beta+(q-2^{-n/T})\Omega_U(T)+\Omega_V(T)$.
On the other hand,
since $T$ is computable,
$\beta+(q-2^{-n/T})\Omega_U(T)$ is an r.e.~real number.
This completes the proof.
\end{proof}

\begin{theorem}\label{(v)->(vii)}
Suppose that $T$ is computable.
For every real number $\alpha\in(0,1)$,
if for every r.e.~$T$-convergent real number $\gamma>0$
there exist an r.e.~real number $\beta\ge 0$ and
a rational number $q>0$ such that
$\alpha=\beta+q\gamma$,
then there exists a universal probability $m$ such that
$\alpha=\sum_{s\in\X}m(s)^{\frac{1}{T}}$.
\end{theorem}

\begin{proof}
First, based on the optimal computer $U$
we define a computer $V\colon\X\to\X$
by the conditions that
(i) $H_V(s)=H(s)+1$ for every $s\in\X$ and
(ii) for every $s\in\X$ and every $n\in\N$,
if $n>H(s)$ then
there exists a unique $p\in\X$ such that
$\abs{p}=n$ and $V(p)=s$.
The existence of such a computer $V$
can be easily shown
using Theorem 3.2 of \cite{C75},
based on the fact that
the set $\{(n,s)\in\N\times\X\mid n>H(s)\}$ is r.e.
and
\begin{equation*}
  \sum_{n>H(s)}2^{-n}
  =
  \sum_{s\in\X}\sum_{n=H(s)+1}^{\infty}2^{-n}
  =
  \sum_{s\in\X} 2^{-H(s)}
  <1,
\end{equation*}
where the first sum is over all
$(n,s)\in\N\times\X$ with $n>H(s)$.
It follows that $V$ is optimal and
\begin{equation}\label{Planck}
  \Omega_V(T)
  =
  \sum_{s\in\X}\sum_{n=H(s)+1}^{\infty}
  2^{-n/T}
  =\frac{1}{2^{1/T}-1}\sum_{s\in\X}2^{-H(s)/T}.
\end{equation}
By Theorem \ref{tomegavt}, we also see that
$\Omega_V(T)$ is an r.e.~$T$-convergent real number.
Thus, by the assumption,
there exist an r.e.~real number $\beta\ge 0$ and
a rational number $q>0$ such that
$\alpha=\beta+q\Omega_V(T)$.
We choose any one rational number $\varepsilon>0$
such that $\varepsilon\le 1-\alpha^T$ and
$\varepsilon^{1/T}<q/(2^{1/T}-1)$.
It follows from \eqref{Planck} that
\begin{equation}\label{core}
\begin{split}
  \alpha
  &=
  \beta
  +
  \frac{q}{2^{1/T}-1}2^{-H(\lambda)/T}
  +
  \left(
    \frac{q}{2^{1/T}-1}-\varepsilon^{1/T}
  \right)
  \sum_{s\neq\lambda}2^{-H(s)/T}\\
  &\hspace*{5mm}
  +
  \sum_{s\neq\lambda}\left(\varepsilon 2^{-H(s)}\right)^{1/T}.
\end{split}
\end{equation}
Let $\gamma$ be the sum of the first, second, and third terms
on the right-hand side of \eqref{core}.
Then, since $T$ is computable,
$\gamma$ is an r.e.~real number.
We define a function $m\colon \X\to (0,\infty)$ by
$m(s)=\gamma^T$ if $s=\lambda$;
$m(s)=\varepsilon 2^{-H(s)}$ otherwise.
Since $\gamma^T<\alpha^T\le 1-\varepsilon$,
we see that
$\sum_{s\in\X} m(s)<\gamma^T+\varepsilon<1$.
Since $T$ is right-computable,
$\gamma^T$ is an r.e.~real number.
Therefore,
since $2^{-H(s)}$ is a lower-computable semi-measure
by Theorem \ref{eup},
$m$ is also a lower-computable semi-measure.
Thus,
since $2^{-H(s)}$ is a universal probability
by Theorem \ref{eup} again
and $\gamma^T>0$,
it is easy to see that $m$ is a universal probability.
On the other hand,
it follows from \eqref{core} that
$\alpha=\sum_{s\in\X}m(s)^{\frac{1}{T}}$.
This completes the proof.
\end{proof}

Theorem \ref{(vii)->(viii)} below is obtained by generalizing
the proofs of Solovay \cite{Sol75} and Theorem 6.4 of
Calude, Hertling, Khoussainov, and Wang \cite{CHKW01}.

\begin{theorem}\label{(vii)->(viii)}
Suppose that $T$ is computable.
For every $\alpha\in(0,1)$,
if there exists a universal probability $m$ such that
$\alpha=\sum_{s\in\X}m(s)^{\frac{1}{T}}$,
then every computable, increasing sequence of rational numbers
which converges to $\alpha$ is $T$-universal.
\end{theorem}

\begin{proof}
Suppose that
$\{a_n\}$ is an arbitrary
computable, increasing sequence of rational numbers
which converges to $\alpha=\sum_{s\in\X}m(s)^{\frac{1}{T}}$.
Since $m$ is a lower-computable semi-measure and $T$ is left-computable,
there exists a total recursive function $f\colon\N\to\N^+$ such that,
for every $n\in\N$, $f(n)<f(n+1)$ and
\begin{equation}\label{upcisrc}
  \sum_{k=0}^{f(n)-1}m(k)^{\frac{1}{T}}\ge a_n.
\end{equation}
Recall here that we identify $\X$ with $\N$.
We then define a total recursive function $g\colon\N\to\N$ by
$g(k)=\min\{n\in\N\mid k\le f(n)\}$.
It follows that 
$g(f(n))=n$ for every $n\in\N$ and
$\lim_{k\to\infty}g(k)=\infty$.

Suppose that
$\{b_n\}$ is an arbitrary $T$-convergent computable,
increasing and converging sequence of rational numbers.
We then choose any one $d\in\N^+$ with
$\sum_{n=0}^{\infty} (b_{n+1}-b_{n})^T\le d$.
We then define a function $r\colon\N\to[0,\infty)$ by
$r(k)=(b_{g(k+1)}-b_{g(k)})^T/d$.
Since $T$ is computable,
$\{b_n\}$ is $T$-convergent, and $g(k+1)=g(k), g(k)+1$, 
we see that $r$ is a lower-computable semi-measure.
Thus,
since $m$ is a universal probability,
there exists $c\in\N^+$ such that,
for every $k\in\N$, $cm(k)\ge r(k)$.
It follows from \eqref{upcisrc} that, for each $n\in\N$,
\begin{equation*}
  (cd)^{\frac{1}{T}}(\alpha-a_n)
  \ge
  d^{\frac{1}{T}}\sum_{k=f(n)}^{\infty}(cm(k))^{\frac{1}{T}}
  \ge
  \sum_{k=f(n)}^{\infty} (b_{g(k+1)}-b_{g(k)})
  =
  \beta-b_{n},
\end{equation*}
where $\beta=\lim_{n\to\infty} b_n$.
This completes the proof.
\end{proof}

Note that,
using Lemma \ref{ks},
we can directly show that
the condition (ii) implies the condition (iii)
in Theorem \ref{partial randomness}
without assuming the computability of $T\in(0,1]$,
as follows.
Theorem \ref{(ii)->(iii)} below holds for an arbitrary
real number $T\in(0,1]$.

\begin{theorem}\label{(ii)->(iii)}
For every r.e.~real number $\alpha$,
if $\alpha$ is Martin-L\"{o}f $T$-random,
then $\alpha$ is $\Omega(T)$-like.
\end{theorem}

\begin{proof}
Suppose that
$\beta$ is an arbitrary r.e.~$T$-convergent real numbers.
Then there is a $T$-convergent computable,
increasing sequence $\{b_n\}$ of rational numbers which
converges to $\beta$.
Since $\{b_n\}$ is $T$-convergent,
without loss of generality
we can assume that
$\sum_{n=0}^{\infty} (b_{n+1}-b_{n})^T\le 1$.
Since $\alpha$ is r.e.~and Martin-L\"{o}f $T$-random
by the assumption,
it follows from Lemma \ref{ks}
that there exist a computable, increasing sequence $\{a_n\}$
of rational numbers and a rational number $q>0$ such that
$a_{n+1}-a_{n}>q(b_{n+1}-b_{n})$ for every $n\in\N$ and
$\alpha=\lim_{n\to\infty}a_{n}$.
It is then easy to see that
$\alpha-a_{n}>q(\beta-b_{n})$ for every $n\in\N$.
Therefore $\alpha$ dominates $\beta$.
This completes the proof.
\end{proof}

\section{Some results on $T$-convergence}
\label{T-convergence}

In this section,
we investigate some properties of the notion of $T$-convergence.
As one of the applications of Theorem \ref{partial randomness},
the following theorem can be obtained first.

\begin{theorem}\label{T-convergent->T-compressible}
Suppose that $T$ is computable.
For every r.e.~real number $\alpha$,
if $\alpha$ is $T$-convergent,
then $\alpha$ is $T$-compressible. 
\end{theorem}

\begin{proof}
Using (vi) $\Rightarrow$ (iv) of Theorem \ref{partial randomness},
we see that
$H(\alpha_n)\le H((\Omega_U(T))_n)+O(1)$
for every r.e.~$T$-convergent real number $\alpha$.
It follows from Theorem \ref{pomgd} (i) that
$\alpha$ is $T$-compressible
for every r.e.~$T$-convergent real number $\alpha$.
\end{proof}

In the case of $T<1$,
the converse of Theorem \ref{T-convergent->T-compressible}
does not hold,
as seen in the following theorem in a sharper form.

\begin{theorem}\label{p2pt}
Suppose that $T$ is computable and $T<1$.
Then there exists an r.e.~real number $\eta$ such that
(i) $\eta$ is weakly Chaitin $T$-random and $T$-compressible,
and
(ii) $\eta$ is not $T$-convergent.
\qed
\end{theorem}

In order to prove Theorem \ref{p2pt},
the following lemma is useful.

\begin{lemma}\label{subsequence}
\hfill
\begin{enumerate}
  \item If $\{a_n\}$ is a $T$-convergent
    increasing sequence of real numbers,
    then every subsequence of the sequence $\{a_n\}$ is
    also $T$-convergent.
  \item Let $\alpha$ be a $T$-convergent r.e.~real number.
    If $\{a_n\}$ is a computable, increasing sequence of
    rational numbers converging to $\alpha$,
    then there exists a subsequence $\{a'_n\}$ of the sequence $\{a_n\}$
    such that
    $\{a'_n\}$ is a $T$-convergent computable,
    increasing sequence of rational numbers converging to $\alpha$.
\end{enumerate}
\end{lemma}

\begin{proof}
(i)
Let $f\colon\N\to\N$ such that $f(n)<f(n+1)$ for all $n\in\N$.
Then,
using repeatedly the inequality $(x+y)^t\le x^t+y^t$
for real numbers $x,y>0$ and $t\in(0,1]$,
we have
\begin{equation*}
  \left(
    a_{f(n+1)}-a_{f(n)}
  \right)^T
  =
  \left[
    \sum_{k=f(n)}^{f(n+1)-1}(a_{k+1}-a_{k})
  \right]^T
  \le
  \sum_{k=f(n)}^{f(n+1)-1}(a_{k+1}-a_{k})^T.
\end{equation*}
It follows that
\begin{equation*}
  \sum_{n=0}^{m}
  \left(
    a_{f(n+1)}-a_{f(n)}
  \right)^T
  \le
  \sum_{k=f(0)}^{f(m+1)-1}(a_{k+1}-a_{k})^T.
\end{equation*}
Since $\{a_n\}$ is $T$-convergent,
we see that the subsequence $\{a_{f(n)}\}$ of $\{a_n\}$
is also $T$-convergent.

(ii)
We choose any one
$T$-convergent computable, increasing sequence $\{b_n\}$ of
rational numbers converging to $\alpha$.
It is then easy to show that
there exist total recursive functions
$g\colon\N\to\N$ and $h\colon\N\to\N$ such that,
for all $n\in\N$,
(i) $g(n)<g(n+1)$, (ii) $h(n)<h(n+1)$, and
(iii) $b_{g(n)}<a_{h(n)}<b_{g(n+1)}$.
It follows from Lemma \ref{subsequence} (i) that
the subsequence $\{b_{g(n)}\}$ of $\{b_n\}$ is $T$-convergent.
Using the inequality $(x+y)^t\le x^t+y^t$
for real numbers $x,y>0$ and $t\in(0,1]$,
we see that
\begin{equation*}
  \left(a_{h(n+1)}-a_{h(n)}\right)^T
  <
  \left(b_{g(n+2)}-b_{g(n)}\right)^T
  \le
  \left(b_{g(n+2)}-b_{g(n+1)}\right)^T
  +\left(b_{g(n+1)}-b_{g(n)}\right)^T.
\end{equation*}
Thus, we see that
the subsequence $\{a_{h(n)}\}$ of $\{a_n\}$ is
a $T$-convergent computable,
increasing sequence of rational numbers converging to $\alpha$.
\end{proof}

The proof Theorem \ref{p2pt} is given as follows.

\begin{proof}[Proof of Theorem \ref{p2pt}]
We choose any one recursive enumeration $p_0,p_1,p_2,\dots$ of
the r.e.~set $\Dom U$,
and define
$\eta$ by
\begin{equation*}
  \eta=\sum_{i=0}^{\infty} \abs{p_i}2^{-\abs{p_i}/T}.
\end{equation*}
Then, since $T$ is computable and $T<1$,
by Theorem 3 of Tadaki \cite{T08CiE}
we see that $\eta$ is an r.e.~real number which
is weakly Chaitin T-random and T-compressible.%
\footnote{
In Theorem 3 of Tadaki \cite{T08CiE},
$\eta$ is furthermore shown to be Chaitin $T$-random, i.e.,
$\lim_{n\to\infty}H(\eta_n)-Tn=\infty$ holds.
}
Since $T$ is computable,
it is easy to show that
there exists a computable, increasing sequence $\{a_n\}$
of rational numbers such that
\begin{equation}\label{sp2pt}
  \sum_{i=0}^{n-1} \abs{p_i}2^{-\abs{p_i}/T}
  < a_{n}
  < \sum_{i=0}^{n} \abs{p_i}2^{-\abs{p_i}/T}
\end{equation}
for all $n\in\N^+$.
Obviously,
$\{a_n\}$ is an increasing sequence of rational numbers
converging to $\eta$.

To show that $\eta$ satisfies the condition (ii) of Theorem \ref{p2pt},
let us assume contrarily that $\eta$ is $T$-convergent.
Then it follows from Lemma \ref{subsequence} (ii) that
there exists a total recursive function $f\colon\N\to\N$ such that
$f(n)<f(n+1)$ for all $n\in\N$, and
$\{a_{f(n)}\}$ is a $T$-convergent computable, increasing sequence of
rational numbers converging to $\eta$.
On the other hand,
since $T$ is computable,
it is easy to show that
there exists a computable, increasing sequence $\{b_n\}$
of rational numbers such that
\begin{equation}\label{s2pt}
  \sum_{i=0}^{f(n)} 2^{-\abs{p_i}/T}
  < b_{n}
  < \sum_{i=0}^{f(n+1)} 2^{-\abs{p_i}/T}
\end{equation}
for all $n\in\N$.
Obviously,
$\{b_n\}$ is an increasing sequence of rational numbers
converging to $\Omega_U(T)$.
Since $U$ is an optimal computer,
using (vi) $\Rightarrow$ (viii) of Theorem \ref{partial randomness},
we see that
there exists $c\in\N^+$ such that
$c(\Omega_U(T)-b_n)\ge \eta-a_{f(n)}$ for all $n\in\N$.
It follows from \eqref{sp2pt} and \eqref{s2pt} that
\begin{equation*}
  c\sum_{i=f(n)+1}^{\infty} 2^{-\abs{p_i}/T}
  >
  \sum_{i=f(n)+1}^{\infty} \abs{p_i}2^{-\abs{p_i}/T}
\end{equation*}
for all $n\in\N^+$.
Therefore, we have
\begin{equation}\label{positivity}
  \sum_{i=f(n)+1}^{\infty} (c-\abs{p_i})2^{-\abs{p_i}/T}>0
\end{equation}
for all $n\in\N^+$.
On the other hand,
it is easy to show that
$\lim_{i\to\infty}\abs{p_i}=\infty$.
Therefore,
since $\lim_{n\to\infty}f(n)=\infty$,
there exists $n_0\in\N^+$ such that,
for all $i\in\N$,
if $i\ge f(n_0)+1$ then $\abs{p_i}\ge c$.
Thus, by setting $n$ to $n_0$ in \eqref{positivity},
we have a contradiction.
This completes the proof.
\end{proof}

Let $T_1$ and $T_2$ be arbitrary computable real numbers
with $0<T_1<T_2<1$,
and let $V$ be an arbitrary optimal computer.
By Theorem \ref{pomgd} (i) and Theorem \ref{T-convergent->T-compressible},
we see that
the r.e.~real number $\Omega_V(T_2)$ is not $T_1$-convergent
and therefore
every computable, increasing sequence $\{a_n\}$ of rational numbers
which converges to $\Omega_V(T_2)$ is not $T_1$-convergent.
Thus,
conversely,
the following question naturally arises:
Is there any computable, increasing sequence of rational numbers
which converges to $\Omega_V(T_1)$ and
which is not $T_2$-convergent~?
We can answer this question affirmatively
in the following form.

\begin{theorem}
Let $T_1$ and $T_2$ be arbitrary computable real numbers
with $0<T_1<T_2<1$.
Then
there exist an optimal computer $V$ and
a computable, increasing sequence $\{a_n\}$ of
rational numbers such that
(i) $\Omega_V(T_1)=\lim_{n\to\infty}a_n$,
(ii) $\{a_n\}$ is $T$-convergent for every $T\in(T_2,\infty)$, and
(iii) $\{a_n\}$ is not $T$-convergent for every $T\in(0,T_2]$.
\end{theorem}

\begin{proof}
First, we choose any one
computable, increasing sequence $\{c_n\}$ of real numbers such that
(i) $\{c_n\}$ converges to a computable real number $\gamma>0$,
(ii) $\{c_n\}$ is $T$-convergent for every $T\in(T_2,\infty)$, and
(iii) $\{c_n\}$ is not $T$-convergent for every $T\in(0,T_2]$.
Such $\{c_n\}$ can be obtained, for example, in the following manner.

Let $\{c_n\}$ be an increasing sequence of real numbers with
\begin{equation*}
  c_n=\sum_{k=1}^{n+1}\left(\frac{1}{k}\right)^{\frac{1}{T_2}}.
\end{equation*}
Since $T_2>0$, we first see that
$\{c_n\}$ is $T$-convergent for every $T\in(T_2,\infty)$, and
$\{c_n\}$ is not $T$-convergent for every $T\in(0,T_2]$.
Since $T_2$ is a computable real number with $0<T_2<1$,
it is easy to see that
$\{c_n\}$ is a computable sequence of real numbers
which converges to a computable real number
$\gamma>0$.
Thus, this sequence $\{c_n\}$ has
the properties (i), (ii), and (iii)
desired above.

We choose any one rational number $r$ with $0<r<1/\gamma$,
and let $\beta=r\gamma$.
Obviously, $\beta$ is a computable real number with $0<\beta<1$.
Let
$b=2^{\frac{1}{T_1}}$.
Then $1<b$.
We can then effectively expand $\beta$ to the base-$b$,
i.e.,
Property \ref{base-b} below holds
for the pair of $\beta$ and $b$.

\begin{property}\label{base-b}
  There exists a total recursive function
  $f\colon\N^+ \to \N$
  such that
  $f(k)\le \lceil b \rceil-1$ for all $k\in\N^+$
  and $\beta=\sum_{k=1}^{\infty}f(k)b^{-k}$.
\end{property}

This can be possible since both $\beta$ and $b$ are computable.
The detail is as follows.
In the case where Property \ref{finiteness} below holds
for the pair of $\beta$ and $b$,
Property \ref{base-b} holds,
obviously.

\begin{property}\label{finiteness}
  There exist $m\in\N^+$ and a function
  $g\colon\{1,2,\dots,m\} \to \N$
  such that
  $g(k)\le \lceil b \rceil-1$ for all $k\in\{1,2,\dots,m\}$
  and $\beta=\sum_{k=1}^{m}g(k)b^{-k}$.
\end{property}

Thus, in what follows, we assume that
Property \ref{finiteness} does not hold.
In this case,
we construct the total recursive function $f\colon\N^+ \to \N$
by calculating $f(1),f(2),f(3),\dots,f(m),\dotsc$
one by one in this order,
based on recursion on stages $m$.
We start with stage $1$ and follow the instructions below.
Note there that the sum $\sum_{k=1}^{m-1}f(k)b^{-k}$ is regarded as $0$
in the case of $m=1$.

At the beginning of stage $m$,
assume that $f(1),f(2),f(3),\dots,f(m-1)$ are calculated already.
We approximate the real number $\beta-\sum_{k=1}^{m-1}f(k)b^{-k}$
and the $\lceil b \rceil -1$ real numbers
\begin{equation*}
  b^{-m},\;2b^{-m},\;\dots,\;
  (\lceil b \rceil -2)b^{-m},\;(\lceil b \rceil -1)b^{-m}
\end{equation*}
by rational numbers with increasing precision.
During the approximation,
if we find $l\in\{0,1,2,\dots,\lceil b \rceil -1\}$
such that
\begin{equation}\label{effecl}
  lb^{-m}<\beta-\sum_{k=1}^{m-1}f(k)b^{-k}<(l+1)b^{-m},
\end{equation}
then we set $f(m):=l$ and begin stage $m+1$.

We can check that our recursion works properly, as follows.
Since $0<\beta<1\le \lceil b \rceil b^{-1}$,
we see that
$0<\beta-\sum_{k=1}^{m-1}f(k)b^{-k}<\lceil b \rceil b^{-m}$
at the beginning of stage $m=1$.
Thus,
in general,
we assume that
$0<\beta-\sum_{k=1}^{m-1}f(k)b^{-k}<\lceil b \rceil b^{-m}$
at the beginning of stage $m$.
Then,
since $\beta$ and $b$ are computable and
Property \ref{finiteness} does not hold,
we can eventually find
$l\in\{0,1,2,\dots,\lceil b \rceil -1\}$ which
satisfies \eqref{effecl}.
Since $b^{-m}\le \lceil b \rceil b^{-(m+1)}$,
we have
$0<\beta-\sum_{k=1}^{m}f(k)b^{-k}<\lceil b \rceil b^{-(m+1)}$
at the beginning of stage $m+1$.

Thus,
Property \ref{base-b} holds in any case.
We choose any one $L\in\N$ with $2^L\ge \lceil b \rceil -1$.
Then
$\sum_{k=1}^{\infty}f(k)2^{-(k+L)}
\le\sum_{k=1}^{\infty}(\lceil b \rceil -1)2^{-(k+L)}
\le 1$.
Hence, by Theorem 3.2 of \cite{C75},
it is easy to show that
there exists a computer $C$ such that
(i) $\#\{\,p\mid \abs{p}=k+L\text{ \& }p\in\Dom C\,\}=f(k)$
for every $k\in\N^+$, and
(ii) $\abs{p}\ge 1+L$ for every $p\in\Dom C$.
We then define a partial function $V\colon\X\to\X$
by the conditions that
(i) $\Dom V=\{\,0p\mid p\in\Dom U\,\}\cup\{\,1p\mid p\in\Dom C\,\}$,
(ii) $V(0p)=U(p)$ for all $p\in\Dom U$, and
(iii) $V(1p)=C(p)$ for all $p\in\Dom C$.
Since $\Dom V$ is a prefix-free set,
it follows that $V$ is a computer.
It is then easy to check that
$H_V(s)\le H_U(s)+1$ for every $s\in\X$.
Therefore, since $U$ is an optimal computer,
$V$ is also an optimal computer.
On the other hand,
we see that
\begin{equation}\label{opluso}
\begin{split}
  \Omega_V(T_1)
  &=\sum_{p\in\Dom U}2^{-(\abs{p}+1)/T_1}
   +
   \sum_{p\in\Dom C}2^{-(\abs{p}+1)/T_1}\\
  &=2^{-\frac{1}{T_1}}\Omega_U(T_1)
    +
    2^{-\frac{L+1}{T_1}}\sum_{k=1}^{\infty}f(k)2^{-k/T_1}\\
  &=2^{-\frac{1}{T_1}}\Omega_U(T_1)
    +
    2^{-\frac{L+1}{T_1}}\beta.
\end{split}
\end{equation}

Since $T_1$ is computable with $0<T_1<1$,
it follows from Theorem \ref{tomegavt} that
there exists a $T_1$-convergent computable,
increasing sequence $\{w_n\}$ of rational numbers which
converges to $\Omega_U(T_1)$.
Then,
since $T_1$ is computable,
it is easy to show that
there exists a computable, increasing sequence $\{a_n\}$ of
rational numbers such that
\begin{equation*}
  \eta w_{n}+\xi c_{n}
  <
  a_n
  <
  \eta w_{n+1}+\xi c_{n+1}
\end{equation*}
for all $n\in\N$,
where $\eta=2^{-\frac{1}{T_1}}$ and
$\xi=2^{-\frac{L+1}{T_1}}r$.
Obviously, by \eqref{opluso} we have
$\lim_{n\to\infty}a_n
=
2^{-\frac{1}{T_1}}\Omega_U(T_1)
+
2^{-\frac{L+1}{T_1}}r\gamma
=
\Omega_V(T_1)$.
Using the inequality $(x+y)^t\le x^t+y^t$
for real numbers $x,y>0$ and $t\in(0,1]$,
we have
\begin{align*}
  (a_{n+1}-a_{n})^T
  &<
  \left[
    (\eta w_{n+2}+\xi c_{n+2})-(\eta w_{n}+\xi c_{n})
  \right]^T \\
  &\le
  \eta^T(w_{n+2}-w_{n})^T+\xi^T(c_{n+2}-c_{n})^T \\
  &\le
  \eta^T(w_{n+2}-w_{n+1})^T+\eta^T(w_{n+1}-w_{n})^T \\
  &\hspace*{5mm}+\xi^T(c_{n+2}-c_{n+1})^T+\xi^T(c_{n+1}-c_{n})^T.
\end{align*}
Thus,
for each $T\in(T_2,\infty)$,
since both $\{w_n\}$ and $\{c_n\}$ are $T$-convergent,
$\{a_n\}$ is also $T$-convergent.
We also have
\begin{align*}
  (c_{n+2}-c_{n+1})^T
  &<
  \left[
    \eta (w_{n+2}-w_{n+1})+\xi (c_{n+2}-c_{n+1})
  \right]^T/\xi^T \\
  &=
  \left[
    (\eta w_{n+2}+\xi c_{n+2})-(\eta w_{n+1}+\xi c_{n+1})
  \right]^T/\xi^T \\
  &<
  (a_{n+2}-a_{n})^T/\xi^T \\
  &\le
  (a_{n+2}-a_{n+1})^T/\xi^T+(a_{n+1}-a_{n})^T/\xi^T.
\end{align*}
Thus,
for each $T\in(0,T_2]$,
since $\{c_n\}$ is not $T$-convergent,
it is easy to see that
$\{a_n\}$ is not $T$-convergent also.
This completes the proof.
\end{proof}

\section{Concluding remarks}
\label{conclusion}

In this paper,
we have generalized
the equivalent characterizations of randomness
for a recursively enumerable real
over the notion of partial randomness,
so that the generalized characterizations
are all equivalent to the weak Chaitin $T$-randomness.
As a stronger notion of partial randomness of a real number $\alpha$,
Tadaki \cite{T99,T02} introduced
the notion of the Chaitin $T$-randomness of $\alpha$,
which is defined as the condition on $\alpha$ that
$\lim_{n\to\infty}H(\alpha_n)-Tn=\infty$.%
\footnote{
The actual separation of the Chaitin $T$-randomness
from the weak Chaitin $T$-randomness is done by
Reimann and Stephan \cite{RS05}.}
Thus,
future work may aim at modifying
our equivalent characterizations of partial randomness
so that they
become
equivalent to the Chaitin $T$-randomness.

\section*{Acknowledgments}

This work was supported
both
by KAKENHI, Grant-in-Aid for Scientific Research (C) (20540134)
and
by SCOPE
(Strategic Information and Communications R\&D Promotion Programme)
from the Ministry of Internal Affairs and Communications of Japan.



\begin{thebibliography}{99}

\bibitem{CHKW01}
C. S. Calude, P. H. Hertling, B. Khoussainov, and Y. Wang,
``Recursively enumerable reals and Chaitin $\Omega$ numbers,''
\emph{Theoret. Comput. Sci}, vol.\ 255, pp.~125--149, 2001.

\bibitem{CST06}
C. S. Calude, L. Staiger, and S. A. Terwijn,
``On partial randomness,''
\emph{Annals of Pure and Applied Logic}, vol.\ 138, pp.~20--30, 2006.

\bibitem{CS06}
C. S. Calude and M. A. Stay,
``Natural halting probabilities, partial randomness,
and zeta functions,''
\emph{Inform. and Comput.}, vol.\ 204, pp.~1718--1739, 2006.

\bibitem{C75}
G. J. Chaitin,
``A theory of program size formally identical to information theory,''
\emph{J. Assoc. Comput. Mach.}, vol.\ 22, pp.~329--340, 1975.

\bibitem{C87b}
G. J. Chaitin,
\emph{Algorithmic Information Theory}.
Cambridge University Press, Cambridge, 1987.

\bibitem{DR07}
R. G. Downey and J. Reimann (2007)
Algorithmic randomness. Scholarpedia, 2(10):2574.
Available at URL:
\url{http://www.scholarpedia.org/article/Algorithmic_randomness}

\bibitem{G74}
P. G\'acs,
``On the symmetry of algorithmic information,''
\emph{Soviet Math. Dokl.}, vol.\ 15, pp.~1477--1480, 1974;
correction, ibid. vol.\ 15, pp.~1480, 1974.

\bibitem{KS01}
A. Ku\v{c}era and T. A. Slaman,
``Randomness and recursive enumerability,''
\emph{SIAM J.\ Comput.}, vol.\ 31, No.\ 1, pp.~199--211, 2001.

\bibitem{L74}
L. A. Levin,
``Laws of information conservation (non-growth) and
aspects of the foundations of probability theory,''
\emph{Problems of Inform. Transmission}, vol.\ 10, pp.~206--210, 1974.

\bibitem{M66}
P. Martin-L\"{o}f,
``The definition of random sequences,''
\emph{Information and Control}, vol.\ 9, pp.~602--619, 1966.

\bibitem{PR89}
M. B. Pour-El and J. I. Richards,
\emph{Computability in Analysis and Physics}.
Perspectives in Mathematical Logic,
Springer-Verlag, Berlin, 1989.

\bibitem{RS05}
J. Reimann and F. Stephan,
\newblock On hierarchies of randomness tests.
\newblock Proceedings of
the 9th Asian Logic Conference,
World Scientific Publishing,
August 16-19, 2005, Novosibirsk, Russia.

\bibitem{Sch73}
C.-P. Schnorr,
``Process complexity and effective random tests,''
\emph{J.\ Comput.\ System Sci.}, vol.\ 7, pp.~376--388, 1973.

\bibitem{Sol75}
R. M. Solovay,
``Draft of a paper (or series of papers) on Chaitin's work ...
done for the most part during the period of Sept.--Dec.\ 1974,''
unpublished manuscript,
IBM Thomas J.\ Watson Research Center, Yorktown Heights, New York,
May 1975, 215 pp.

\bibitem{T99}
K. Tadaki,
\newblock Algorithmic information theory and fractal sets.
\newblock Proceedings of
1999 Workshop on Information-Based Induction Sciences (IBIS'99),
pp.~105--110,
August 26-27, 1999, Syuzenji, Shizuoka, Japan.
In Japanese.

\bibitem{T02}
K. Tadaki,
``A generalization of Chaitin's halting probability $\Omega$ and
halting self-similar sets,''
\emph{Hokkaido Math.\ J.}, vol.\ 31, pp.~219--253,
2002.
Electronic Version Available:
\url{http://arxiv.org/abs/nlin/0212001v1}

\bibitem{T06}
K. Tadaki,
``An extension of Chaitin's halting probability $\Omega$ to
a measurement operator in an infinite dimensional quantum system,''
\emph{Math.\ Log.\ Quart.}, vol.\ 52, pp.~419--438,
2006.

\bibitem{T08CiE}
K. Tadaki,
\newblock A statistical mechanical interpretation of
algorithmic information theory.
\newblock To appear in the
Proceedings of Computability in Europe 2008 (CiE 2008),
June 15-20, 2008, University of Athens, Greece.
Extended and Electronic Version Available:
\url{http://arxiv.org/abs/0801.4194v1}

\bibitem{T08ISIT}
K. Tadaki,
\newblock The Tsallis entropy and the Shannon entropy of
a universal probability.
\newblock To appear in the Proceedings of
the 2008 IEEE International Symposium on Information Theory
(ISIT2008), July 6-11, 2008, Toronto, Canada.
Electronic Version Available:
\url{http://arxiv.org/abs/0805.0154v1}

\bibitem{W00}
K. Weihrauch,
\emph{Computable Analysis}.
Springer-Verlag, Berlin, 2000.

\bibitem{ZL70}
A. K. Zvonkin and L. A. Levin,
``The complexity of finite objects and the development of the concepts of
information and randomness by means of the theory of algorithms,''
\emph{Russian Math. Surveys}, vol.\ 25, no.\ 6, pp.~83--124, 1970.

\end{thebibliography}
\end{document}